\documentclass[prl,twocolumn,superscriptaddress,preprintnumbers,amsmath,amssymb]{revtex4}
\usepackage[dvips]{graphicx}
\usepackage{dcolumn}
\usepackage{color}
\usepackage{bm}

\begin{document}

\title{Quantum jump approach for work and dissipation in a two-level system}

\author{F. W. J. Hekking}
\affiliation{LPMMC, Universite Joseph Fourier and CNRS, 25 Avenue des Martyrs, BP 166, 38042 Grenoble, France}

\author{J. P. Pekola}
\affiliation{Low Temperature Laboratory (OVLL), Aalto University School of Science, P.O.
Box 13500, 00076 Aalto, Finland}

\date{\today}

\begin{abstract}
We apply the quantum jump approach to address the statistics of work in a driven
two-level system coupled to a heat bath. We demonstrate how this question can be analyzed
by counting photons absorbed and emitted by the environment in repeated experiments. We
find that the common non-equilibrium fluctuation relations are satisfied identically.
The usual fluctuation-dissipation theorem for linear response applies for
weak dissipation and/or weak drive. We point out qualitative differences between the
classical and quantum regimes.
\end{abstract}

\maketitle

The Quantum Jump (QJ) method, also called Monte Carlo wave function technique, was developed in the early 90s
\cite{dalibard92,gardiner92,carmichael93,plenio98}. This development followed a series of
experiments performed in the mid 80s that reported the observation of QJs in
ions \cite{nagourney86,sauter86,bergquist86}, in conjunction with theoretical
work concerning the nature of these jumps
\cite{cook85,cohen_tannoudji86,javanainen86}. Subsequent experiments in quantum optics
\cite{gleyzes07} have probed QJs associated with the birth and death of photons in a
cavity. It was shown that averaging of many individual single photon quantum trajectories
is equivalent to solving the relevant master equation \cite{carmichael93,raimond06}. More
recently, the QJ method has also been used to address issues related to measurements on
quantum systems~\cite{korotkov99,gambetta08,wisemen12}.

\begin{figure}
   \includegraphics[width=8.5cm]{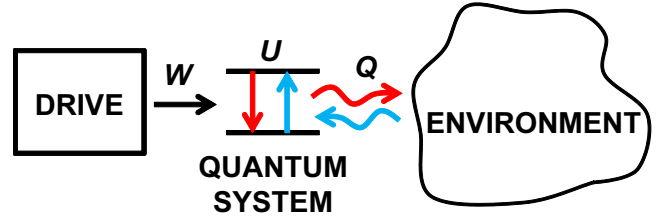}
    \caption{A quantum two-level system (centre) driven by an external force
    (left) and coupled to an environment (right).}
    \label{Fig1}
\end{figure}

In this paper we propose to use the QJ method as an efficient means to discuss the
problem of determining the statistics of work in driven quantum systems with dissipation,
currently a topic of intense discussion \cite{esposito09,campisi11}. In particular,
unlike for classical systems \cite{jarzynski97,crooks99}, the full statistics of work and
the resulting nonequilibrium fluctuation relations are still not well established for
quantum systems.  We approach the problem by constructing quantum trajectories based on
the QJ method, and demonstrate the validity of non-equilibrium fluctuation relations in a
driven two-level system (qubit) coupled to a dissipative environment. Using the same
technique we discuss the two lowest moments of work in driven evolution. Figure
\ref{Fig1} presents schematically the set-up we consider. A two-level quantum system is
driven by a classical source exerting work $W$ on it. The system is also coupled to a
thermal bath, with which it can exchange heat $Q$. The dynamics of the two-level system
is determined by the combined action of the source and the environment.

We start by considering the driven two-level system in the absence of dissipation, described by the Hamiltonian
\begin{equation} \label{q1}
H_S =\hbar \omega_0 a^\dagger a + \lambda(t)(a^\dagger +a),
\end{equation}
where $a^\dagger =|e\rangle \langle g|$ and $a =|g\rangle \langle e|$ are the creation
and annihilation operators in the ground $|g\rangle$ - excited $|e\rangle$ state basis of
the undriven system, $\hbar \omega_0 = E_e - E_g$ is the energy separation of the two
levels, and $\lambda(t)$ is the drive signal of the source. A normalized quantum state
$|\psi(t)\rangle$ describing this system at arbitrary time $t$ can always be written as a
superposition of the states $|g\rangle$ and $|e\rangle$: $|\psi(t)\rangle = a(t)|g\rangle
+ b(t)|e\rangle$ with $|a(t)|^2 +|b(t)|^2= 1$. The infinitesimal time evolution of such a
state is governed by the equation
\begin{equation}
|\psi(t+\Delta t)\rangle = [1- i \Delta t H_S/\hbar]|\psi(t)\rangle, \label{evolution}
\end{equation}
which conserves the normalization.

Next consider the driven two-level system coupled to a bath with which it can exchange
photons of frequency $\omega_\mu$. For definiteness, we assume the system-bath coupling
Hamiltonian $H_C$ to have the linear form
\begin{equation}
H_C = \sum _\mu c_\mu a^\dagger b_\mu  + c_\mu^*
b_\mu^\dagger a.
\end{equation}
Suppose that at time $t$ the total system is in a state
\begin{equation}
|\psi(t)\rangle = [a(t)|g\rangle + b(t) |e\rangle] \otimes |0\rangle,
\end{equation}
{\em i.e.}, the two-level system is in a generic superposition state, and the bath in a
state without any excess photons. At a slightly later time $t+\Delta t$ we have
\begin{equation}
|\psi(t+\Delta t)\rangle = |\psi^{(0)}(t+\Delta t)\rangle + |\psi^{(1)}(t+\Delta
t)\rangle + |\psi^{(-1)}(t+\Delta t)\rangle, \label{state}
\end{equation}
where
\begin{eqnarray}
|\psi^{(0)}(t+\Delta t)\rangle = [a(t+\Delta t)|g\rangle + b(t+\Delta t) |e\rangle] \otimes |0\rangle, \\
|\psi^{(1)}(t+\Delta t)\rangle = \sum _\mu \beta_{\mu,+} |g\rangle \otimes |n_\mu +1 \rangle, \\
|\psi^{(-1)}(t+\Delta t)\rangle = \sum _\mu \beta_{\mu,-}  |e\rangle \otimes |n_\mu -1
\rangle.
\end{eqnarray}
We assumed $\Delta t$ to be short enough that at most one photon is exchanged with the
bath. The various components therefore involve only $0$, $1$ or $-1$ excess photons
$\mu$. The amplitudes $\beta_{\mu, \pm}$ can be obtained using standard time-dependent
perturbation theory with respect to $H_C$; to the lowest order one finds
\begin{equation}
\sum_\mu |\beta_{\mu,+}|^2 =  |b(t)|^2\Gamma_\downarrow \Delta t \mbox{, }
\sum_\mu |\beta_{\mu,-}|^2 =
 |a(t)|^2\Gamma_\uparrow \Delta t,
\end{equation}
where
\begin{eqnarray}
\Gamma_\downarrow &=&
\frac{2 \pi}{\hbar} \sum _\mu (n_\mu +1)|c_\mu|^2
\delta
(\hbar \omega_0- \hbar \omega_\mu), \\
\Gamma_\uparrow &=& \frac{2 \pi}{\hbar}  \sum_\mu n_\mu |c_\mu|^2
\delta (\hbar \omega_0- \hbar \omega_\mu)
\end{eqnarray}
are the photon emission and absorption rate, respectively. Here we assume that the time
step $\Delta t$ is short compared to the relevant time scale of the dynamics of the
two-level system, yet long compared to the bath's correlation time so that energy
conservation be accurate~\cite{footnote}. Note that $\Gamma_\uparrow/\Gamma_\downarrow = e^{-\beta \hbar
\omega_0}$ (detailed balance), provided the bath remains in thermal equilibrium at all
times, such that $n_\mu = (e^{\beta \hbar \omega_\mu} -1)^{-1}$.

In order for $|\psi(t+\Delta t)\rangle$, Eq.~(\ref{state}), to be normalized, we have to
impose $\langle \psi^{(0)}(t+\Delta t)|\psi^{(0)}(t+\Delta t)\rangle = 1 - \Delta p$
where
\begin{equation}
\Delta p = \Delta t[|a(t)|^2\Gamma_\uparrow  + |b(t)|^2\Gamma_\downarrow].
\end{equation}
This can be achieved by modifying the standard time evolution
into a non-hermitian one, replacing in (\ref{evolution}) the Hamiltonian $H_S$ by
\begin{equation}
H =  H_S - i \hbar \Gamma_\downarrow |e\rangle\langle e|/2 - i \hbar \Gamma_\uparrow |g
\rangle \langle g|/2.
\end{equation}

We are now in a position to define the QJ
procedure. Let at time $t$ the system be in the normalized state $|\psi(t)\rangle$. If no
photon exchange occurs during the time interval $\Delta t$, it will be in a state
\begin{equation}
|\psi^{(0)}(t+\Delta t)\rangle =  [1 - i \Delta t H/\hbar]|\psi(t)\rangle
\end{equation}
at time $t+\Delta t$, with norm $1- \Delta p$. Hence the normalized state $|\psi(t+\Delta
t)\rangle = |\psi^{(0)}(t+\Delta t)\rangle /\sqrt{1-\Delta p}$. Should a photon exchange
(a QJ) occur during $\Delta t$, the normalized state will be either $|\psi(t+\Delta
t)\rangle = |e\rangle $ or $|\psi(t+\Delta t)\rangle = |g\rangle $, depending on whether
the photon was absorbed or emitted by the two-level system. The Monte Carlo procedure
consists of choosing a random number $\epsilon$ between zero and one. If $\epsilon >
\Delta p$, no QJ occurs, and we take $|\psi(t+\Delta t)\rangle = |\psi^{(0)}(t+\Delta
t)\rangle /\sqrt{1-\Delta p}$. If $\epsilon < \Delta p$, a photon is either emitted with
probability $|b(t)|^2\Gamma_\downarrow/[|a(t)|^2\Gamma_\uparrow  +
|b(t)|^2\Gamma_\downarrow]$ (state $|\psi(t+\Delta t)\rangle = |g\rangle $)  or absorbed
with probability $|a(t)|^2\Gamma_\uparrow/[|a(t)|^2\Gamma_\uparrow  +
|b(t)|^2\Gamma_\downarrow]$ (state $|\psi(t+\Delta t)\rangle = |e\rangle $). It is easy
to show (see supplemental material) that this procedure is equivalent to the analysis of
the usual master equation for the partial density matrix, defined as the average of
$|\psi(t)\rangle\langle \psi(t)|$ over the bath degrees of freedom~\cite{molmer93}.
Moreover, this procedure shows that the environment not only induces QJs, but also
influences the evolution of the system in between such jumps, where the dynamics of the
amplitudes $a$ and $b$ in the interaction representation is governed by
\begin{eqnarray}
i \hbar \dot{a} = e^{-i\omega_0 t}\lambda(t) b + i \hbar \Delta \Gamma |b(t)|^2 a(t)/2, \label{dyna} \\
i \hbar \dot{b} = e^{i\omega_0 t}\lambda(t) a - i\hbar \Delta \Gamma |a(t)|^2 b(t)/2 \label{dynb},
\end{eqnarray}
with $\Delta \Gamma = \Gamma_\downarrow - \Gamma_\uparrow$.

\begin{figure}
  \includegraphics[width=8.5cm]{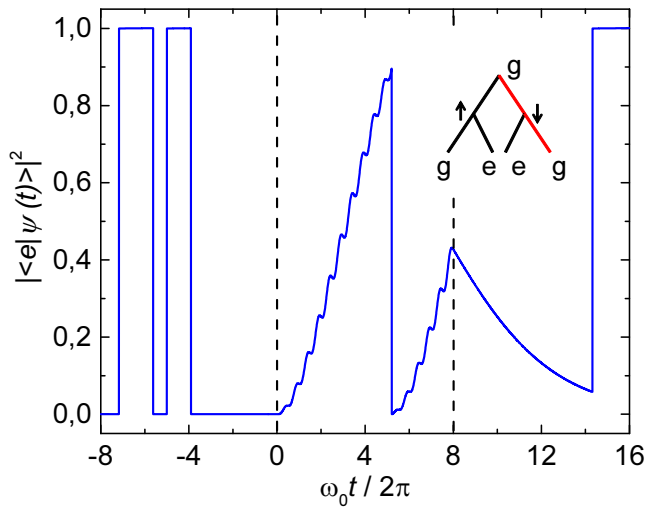}
    \caption{An example of a quantum jump simulation. The environment
temperature is $\beta \hbar \omega_0 = 1.0$, the amplitude of the harmonic drive
(frequency $\omega = \omega_0$) is $\lambda_0 = 0.1\hbar\omega_0$ and lasts $8$ cycles,
the relaxation rate is given by $\Gamma_\downarrow = 0.1\omega_0$.}
    \label{Fig2}
\end{figure}

Figure \ref{Fig2} is a numerical example of the evolution of the excited state population
$|\langle e|\psi(t)\rangle|^2$, obtained using the QJ procedure. At times $t < 0$ the
system is not driven, and it jumps between the two eigenstates $|e\rangle$ and
$|g\rangle$ stochastically, governed by the rates $\Gamma_\downarrow$ and
$\Gamma_\uparrow$, and the instantaneous populations. In the time interval $0\le \omega_0
t/2\pi< 8$, the system is driven resonantly by the force $\lambda(t)=\lambda_0
\sin(\omega_0t)$. Within this interval, it makes one QJ down $|e\rangle \rightarrow
|g\rangle$ in this particular realization. Finally, at times $\omega_0 t/2\pi \ge 8$, the
drive is absent again, and the collapse (jump to $|e\rangle$) tells that the system was measured to be
in the ground state at the end of the drive.

We next demonstrate that detecting the photons emitted and absorbed by calorimetry of the
environment serves as a traditional projective measurement. Suppose that the system's
wave function reads $|\psi(\mathcal T)\rangle = a(\mathcal T)|g\rangle + b(\mathcal
T)|e\rangle$ at the end of the driving period. The evolution of the amplitudes at times
$t\ge\mathcal T$ is governed by Eqs.~(\ref{dyna}), (\ref{dynb}) with $\lambda = 0$, until
the first "guardian" photon is exchanged. During this quiet period, the excited state
population $p_e(t)=|b(t)|^2$ thus evolves as $\dot p_e =-\Delta \Gamma p_e(1-p_e)$. Hence
\begin{equation} \label{q3}
p_e(t)=(1+r e^{\Delta \Gamma (t-{\cal T})})^{-1},
\end{equation}
where $r=(1-p_e(\mathcal T))/p_e(\mathcal T)$. We can then evaluate the probability
$P_{\rm E}$ that the system is found to be in the excited state, indicated by the
absorption of the "guardian" photon by the environment (as opposed to being emitted by
the environment). Indeed,
\begin{equation} \label{q4}
P_{\rm E}=\int_{\mathcal T}^\infty \Delta t\Gamma_\downarrow p_e(t) e^{-\int_{\mathcal
T}^t \Delta t'[\Gamma_\uparrow (1-p_e(t'))+\Gamma_\downarrow p_e(t')]},
\end{equation}
and integrating Eq. \eqref{q4} after substitution of $p_e(t)$ from Eq. \eqref{q3} yields
$P_{\rm E}=p_e(\mathcal T)$. This intuitive result holds irrespective of the values of
$\Gamma_\uparrow$ and $\Gamma_\downarrow$, {\em i.e.}, it is valid at any temperature of
the environment and independent of the strength and type of the coupling. This is in
accordance with the result of a projective measurement of the state of a quantum system.

Based on the interpretation of traces as that in Fig.~\ref{Fig2}, we obtain the work in
each realization as follows. We make two measurements, in the spirit of the
two-measurement protocol that was formerly applied to an isolated driven system
\cite{kurchan00,campisi11}: one measurement before the driving period, and another one
after it. These measurements are done by the detection of the last photon
emitted/absorbed by the system to the environment before the drive and of the first
photon after the drive. This can be realized in practice calorimetrically as proposed in
\cite{jp12}. In the example of Fig. \ref{Fig2}: (i) the first measurement indicates that
the initial state of the system is $|g\rangle$ (internal energy $U_i=E_g$), since the
last photon before the application of the force (at $\omega_0 t/2\pi \simeq -4$) was
emission by the system, and (ii) the second measurement shows that the final state of the
system was likewise $|g\rangle$ (internal energy is $U_f= E_g$), since the first photon
after the drive (at $\omega_0 t/2\pi \simeq 14$) was absorption by the system. These two
measurements thus tell that the force has not changed the system internal energy $U$,
i.e., $\Delta U = U_f - U_i = 0$ for this particular realization. (The other possible
outcomes would have been $\Delta U=\pm \hbar\omega_0$.) During the drive, heat is
released to or taken from the environment by the QJ events. Again, in the example of Fig.
\ref{Fig2}, the one photon emitted by the system is equivalent of assigning heat
$Q=+\hbar \omega_0$ released to the environment. The work done by the source is then
$W=\Delta U+Q$, and equals $\hbar\omega_0$ for this realization. Our ultimate task is
then to find the distribution of $W$ in repeated experiments, and to assess the
fluctuation relations and the various moments of $W$.

We proceed by presenting a systematic method to analyze the statistics of work and heat
under the force protocol $\lambda(t)$ from the initial time $0$ to the final time ${\cal
T}$. At time $t=0$, the system is supposed to be equilibrated by the heat bath. As a
result it will occupy the ground state with probability $p_g = (1 + e^{-\beta \hbar
\omega_0})^{-1}$ and the excited state with probability $p_e = e^{- \beta \hbar \omega_0}
p_g$. Therefore, to obtain averages involving $W$ under many repetitions of the protocol
$\lambda(t)$, two cases should be distinguished. One corresponds to the case where the
protocol is run on the ground state, the other to the case where the protocol is run on
the excited state. For both cases, the set of possible quantum trajectories can be
represented with the help of a Cayley tree. The inset of Fig.~\ref{Fig2} shows the Cayley
tree corresponding to all possible quantum trajectories starting from the ground state
and undergoing one QJ during the driving period ${\cal T}$. Specifically, the trajectory
in red is the one realized during the simulation shown in Fig.~\ref{Fig2}.
The
probability $P$ for such a single photon trajectory starting in the ground state is given by
\begin{equation}
P = p_g \Gamma_\downarrow \int _{0}^{\cal T} dt |a_g({\cal T},t)|^2 e^{-\pi_{g}({\cal
T},t)} |b_g(t,0)|^2 e^{-\pi_{g}(t,0)}.
\end{equation}
Here  $b_g(t,0)$ denotes the probability amplitude $b$ at time $t$ with the ground state
$b(0) = 0$ as initial condition at $t=0$. Similarly, $a_g({\cal T},t)$ denotes the
probability amplitude $a$ at time ${\cal T}$ with the ground state $a(t) = 1$ as initial
condition at time $t$. Here $a$ and $b$ are found by solving Eqs.~(\ref{dyna}) and
(\ref{dynb}). The probability that no photon is exchanged with the bath is given by the
Poisson factor $e^{-\pi_{g}}$, where we defined
\begin{equation}
\pi_{g,e}(t_2,t_1) = \int _{t_1}^{t_2} dt [\Gamma_\uparrow |a_{g,e}(t,t_1)|^2 +
\Gamma_\downarrow |b_{g,e}(t,t_1)|^2]. \label{pi_eg}
\end{equation}
The total probability $P_1$ for a one-photon process to occur under the action of the
drive is found by summing over all trajectories for this tree and for the one
corresponding to the initial excited state.

\begin{figure}
   \includegraphics[width=8.5cm]{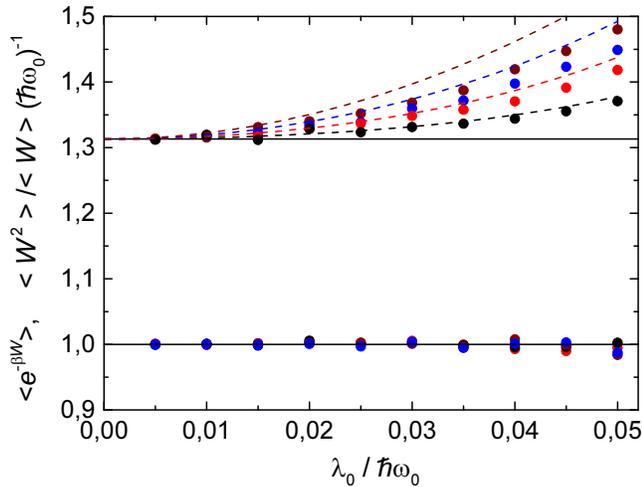}
    \caption{Results of numerical simulations and analytic approximations.
The ratio of the two lowest moments of work and the JE average are shown as functions of
the driving amplitude. The symbols are numerical simulations and the lines are
analytical approximations, the colored ones are for the perturbative in
$\Gamma_{\downarrow,\uparrow}$ calculation. Here, $\beta\hbar\omega_0 =2.0$, the resonant
($\omega=\omega_0$) drive lasts $10$ cycles, and the various data sets are for
$\Gamma_\downarrow = (0,) 0.005,0.01,0.015, 0.02$ from bottom to top.}
    \label{Fig3}
\end{figure}

The calculation of averages involving the quantity $W$ is now immediate. For example,
along the trajectory analyzed above we have $W = \hbar \omega_0$. This trajectory thus
contributes to $\langle W^k \rangle$ as $(\hbar \omega_0)^k P/P_1$ and to $\langle e^{-
\beta W} \rangle$ as $e^{ - \beta \hbar \omega_0} P/P_1$. The other trajectories can be
analyzed similarly; the extension to Cayley trees corresponding to arbitrary $n$-photon
processes is straightforward. In Fig.~\ref{Fig3} we show the results of calculations of
the ratio of the two lowest moments of $W$ as well as of the quantity $\langle e^{-\beta
W}\rangle$. The points are obtained with QJ simulations, the solid lines correspond to a
perturbative solution of Eqs.~(\ref{dyna}) and (\ref{dynb}) for weak dissipation (see
supplemental material). In the linear response limit $\lambda_0 \to 0$, we find that the
ratio $\langle W^2 \rangle/\hbar \omega_0 \langle W \rangle \to \coth (\beta \hbar
\omega_0/2) \simeq 1.31$: the usual fluctuation-dissipation result. As $\lambda_0$ is
increased, deviations are found from linear response that are more important for stronger
dissipation; perturbation theory breaks down at relatively low drive amplitudes.

We now turn to the results for the quantity $\langle e^{-\beta W}\rangle$. The
simulations show that within the numerical accuracy this quantity equals 1 for the
parameter range studied here, in agreement with the celebrated Jarzynski equality (JE)
$\langle e^{-\beta W}\rangle =1$~\cite{jarzynski97}. (Since the drive lasts over an integer number
of periods, the free-energy difference between the initial and final points vanishes,
and the right side of this equation is indeed expected to be equal to unity.) Analyzing the Cayley tree
trajectories systematically, one can demonstrate the validity of the JE for the
dissipative driven two-level system studied here (see supplemental material). The proof
is based on the fact that the quantity $\langle e^{-\beta W}\rangle$ is equal to the
(normalized) total probability for all the trajectories under the reverse protocol
$\lambda_R(t) = \lambda({\cal T}-t)$, provided the rates $\Gamma_\uparrow$, $\Gamma_\downarrow$, as well as the
probabilities $p_e$ and $p_g$, satisfy detailed balance. 
We like to emphasize that on one hand this proof, based on reversed trajectories, is analogous to the early one by 
Crooks for a classical two-state system obeying detailed balance for transition rates \cite{crooks98}. Yet the classical dynamics, 
presenting definite alternating transitions between the two states, differs from the quantum evolution involving superposition states, leading to the branching of the trajectories shown by the Cayley trees.

\begin{figure}
   \includegraphics[width=8.5cm]{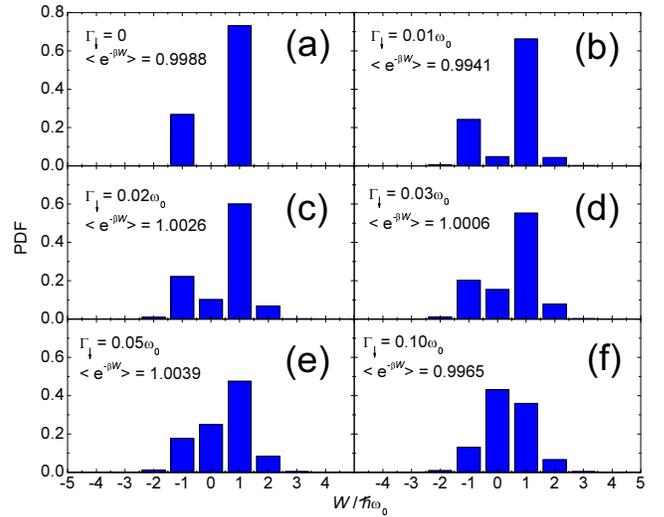}
    \caption{Numerically calculated work distributions under
    the influence of the $\pi$-pulse. The system, initially in thermal equilibrium,
   is driven at the resonance angular frequency $\omega_0$ over 10 periods, with
    amplitude $\lambda_0 = 0.05\hbar\omega_0$. The bath temperature is given by $\beta\hbar\omega_0 =1$,
    and we employed $10^5$ repetitions for each value of $\Gamma_\downarrow$. (a) presents an isolated qubit, whereas (b) - (f) correspond to open systems with increasing relaxation.}
    \label{Fig4}
\end{figure}

A two-level system driven sinusoidally over a time $\mathcal T =\pi /\lambda _0$ at
angular frequency $\omega_0$ undergoes a so-called $\pi$-pulse,
driving it into the excited state if it was initially in the ground state, and vice
versa. Figure \ref{Fig4} shows the corresponding work distribution calculated for various
rates of relaxation. Initially the system is in thermal equilibrium. Figure \ref{Fig4}a
shows the probability distribution function (PDF) for vanishing relaxation rate. In this
case the work has two possible values: $-\hbar\omega_0$ with probability $p_e$, and
$+\hbar\omega_0$ with probability $p_g$. Upon increasing the relaxation rate in (b)-(f),
the PDF evolves from the "bimodal" one into a more bell-shaped distribution. For all
values of relaxation, the JE is satisfied within the numerical error; the values obtained
by $10^5$ repetitions in each case are indicated in the corresponding panel.

A natural realization of the presented scheme is a superconducting phase qubit
\cite{martinis02,claudon04} coupled inductively to a dissipative element, whose
temperature can be monitored in real time in order to perform a calorimetric measurement
\cite{jp12}. Specifically, one may use a current-biased SQUID, yielding a two-level
system with a typical level spacing of the order of $\hbar\omega_0/k_B \sim $ 1 K. The
rates are given by $\Gamma_\downarrow = g^2 S(+\omega_0)$ and $\Gamma_\uparrow=g^2
S(-\omega_0)$. The coupling $g$ is proportional to the mutual inductance between the
SQUID loop and the dissipative element, {\em i.e.}, it is determined by the geometry of
the set-up. The noise spectral function $S(\pm\omega_0)$ of the resistive element is
taken at angular frequency $\pm\omega_0$. For thermal
noise, detailed balance between the $\downarrow,\uparrow$ rates is obeyed.

In summary, we have analyzed work in a disipative two-level quantum system using the
quantum jump approach. The common fluctuation theorem (JE) is shown to be valid, and we
obtain the moments of work distribution in linear response and beyond. As an
illustration, we apply the method to a qubit driven by a $\pi$-pulse and we demonstrate
that the model can be realized for instance as a superconducting phase qubit.

We thank S. Gasparinetti,  T. Ala-Nissila, A. Shnirman, and P. Solinas for discussions.
The work has been supported partially by the Academy of Finland through its LTQ (project no. 250280) CoE grant, the AScI
visiting professor program at Aalto University, European Union FP7 project INFERNOS (grant agreement 308850), and Institut
universitaire de France.

\begin{widetext}
\section{Appendices}

\subsection{Equivalence of quantum jump approach and Master equation}

Let $\rho(t)$ be the density matrix of the complete system defined as $|\psi(t)\rangle
\langle \psi(t)|$, and let $\sigma(t)$ be the reduced density matrix of the two-level
system obtained by performing a partial trace of $\rho(t)$ over the bath degrees of
freedom. With the help of Eqs.~(5) -- (9) from the main text, it is straightforward to
perform the partial trace of $\rho(t+dt)$. Using in addition Eqs.~(14) and (15) from the
main text, we obtain
\begin{equation}
\sigma (t+dt) = \sigma(t) - i H(t) \sigma(t) dt/\hbar + i \sigma(t) H^\dagger (t)
dt/\hbar + \sigma_{ee}(t) \Gamma_\downarrow |g\rangle \langle g|dt +
\sigma_{gg}(t)\Gamma_\uparrow |e\rangle \langle e|dt.
\end{equation}
Here we used the fact that $|a(t)|^2 = \sigma_{gg}(t) \equiv \langle g |\sigma (t)
|g\rangle$ and $|b(t)|^2 = \sigma_{ee}(t) \equiv \langle e |\sigma(t) |e\rangle$. In the
interaction representation with respect to the undriven two-level system, this can be
written as
\begin{equation}
d \sigma_I/dt = - i H_{1,I}(t) \sigma_I(t)/\hbar + i \sigma_I(t) H^\dagger_{1,I} (t)
/\hbar + \sigma_{I,ee}(t) \Gamma_\downarrow |g\rangle \langle g| +
\sigma_{I,gg}(t)\Gamma_\uparrow |e\rangle \langle e|, \label{sigI}
\end{equation}
where the non-Hermitian time-dependent Hamiltonian $H_{1,I}(t)$ is given by
\begin{equation}
H_{I,1}(t) = \lambda(t) [e^{i \omega_0 t}|e\rangle\langle g| + e^{-i\omega_0 t}|g\rangle
\langle e|] - i\hbar/2 [\Gamma_\downarrow |e\rangle\langle e| + \Gamma_\uparrow
|g\rangle\langle g|].
\end{equation}
The matrix elements of (\ref{sigI}) yield the standard Bloch-Redfield master equation for
a driven two-level system in the presence of dissipation,
\begin{eqnarray} \label{ht2}
&& \dot \sigma_{I,gg} = -\frac{2\lambda(t)}{\hbar}\Im {\rm
m}(\sigma_{I,ge}e^{i\omega_0t}) -\Gamma_\Sigma \sigma_{I,gg}+\Gamma_\downarrow, \nonumber
\\&& \dot\sigma_{I,ge} =
\frac{i\lambda(t)}{\hbar}e^{-i\omega_0t}(2\sigma_{I,gg}-1)-\frac{1}{2}\Gamma _\Sigma
\sigma_{I,ge},
\end{eqnarray}
where we defined $\Gamma _\Sigma = \Gamma_\uparrow +\Gamma_\downarrow$.

\subsection{Perturbative analysis of Cayley trees for weak dissipation}

By construction, the Cayley trees allow for a systematic analysis of the contribution to
the statistics of work of quantum trajectories involving processes during which $n=0,1,2,
...$ photons are exchanged with the environment. To illustrate this, we consider the
limit of weak dissipation taking into account contributions to linear order in
$\Gamma_\uparrow$ and $\Gamma_\downarrow$. To this order, at most one photon is exchanged
with the environment, and we need to consider the Cayley trees shown in Fig.~\ref{ctree}a
and b.
\begin{figure}
  \includegraphics[width=10cm]{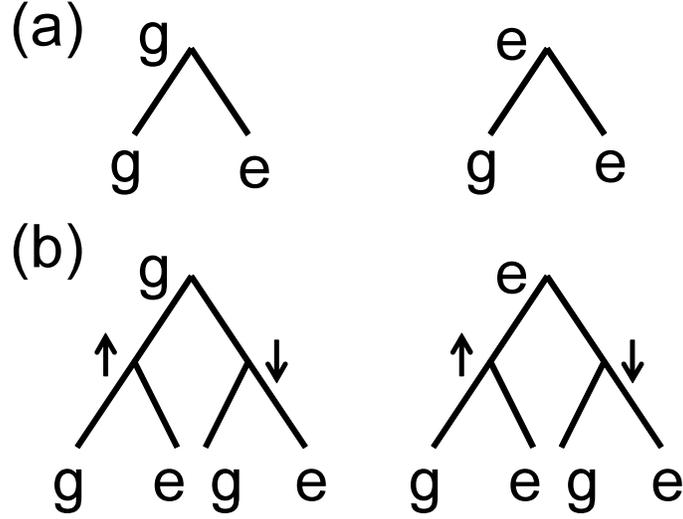}
    \caption{Cayley trees corresponding to (a) zero-photon and (b)
    one-photon exchange between the two-level system and the environment.}
    \label{ctree}
\end{figure}

We start by analyzing the Cayley trees shown in Fig.~\ref{ctree}a. Starting from the
ground state $g$ with probability $p_g$ at the initial time $t=0$, two trajectories are
possible under the drive $\lambda(t)$: the system either stays in the ground state $g$ or
evolves to the excited state $e$ at the final time $t = {\cal T}$ (left panel).
Similarly, when starting from the excited state $e$ with probability $p_e$, the system
either stays in $e$ or evolves to the ground state $g$ (right panel). The four possible
trajectories correspond to the total probability
\begin{equation}
P_0 = p_g e^{-\pi_g({\cal T},0)}[|a_g({\cal T},0)|^2 +|b_g({\cal T},0)|^2 ] + p_e
e^{-\pi_e({\cal T},0)}[|a_e({\cal T},0)|^2 +|b_e({\cal T},0)|^2 ] = p_g e^{-\pi_g({\cal
T},0)} + p_e e^{-\pi_e({\cal T},0)},
\end{equation}
where $\pi_{g,e}$ is defined by Eq.~(20) in the main text. As explained in the main text,
even when no actual photon exchange occurs with the environment, the time evolution is
affected by the environment. We solve Eqs.~(15) and (16) of the main text perturbatively
for a resonant drive, $\lambda(t) = \lambda_0 \sin \omega_0 t$, and find to first order
in $\Delta \Gamma$
\begin{eqnarray}
&&|a_g(t,t_i)|^2 = 1 - |b_g(t,t_i)|^2 = \cos^2 \lambda_0 (t-t_i)/2\hbar + \frac{\hbar
\Delta \Gamma}{\lambda_0} \cos \lambda_0 (t-t_i)/2\hbar \sin^3 \lambda_0 (t-t_i)/2\hbar, \label{perag}\\
&&|a_e(t,t_i)|^2 = 1 - |b_e(t,t_i)|^2 = \sin^2 \lambda_0 (t-t_i)/2\hbar + \frac{\hbar
\Delta \Gamma}{\lambda_0} \cos \lambda_0 (t-t_i)/2\hbar \sin^3 \lambda_0 (t-t_i)/2\hbar
\label{perae}.
\end{eqnarray}
As a result
\begin{equation}
\pi_{g,e}({\cal T},0)\simeq \Gamma_\Sigma {\cal T}/2 \mp \frac{\hbar \Delta \Gamma}{2
\lambda_0} \sin \lambda_0 {\cal T}/\hbar, \label{perpi}
\end{equation}
hence $P_0 \simeq 1 - \Gamma_\Sigma {\cal T}/2 + (p_g-p_e)(\hbar \Delta
\Gamma/2\lambda_0) \sin \lambda_0 {\cal T}/\hbar$.

We are now in a position to calculate moments $\langle W^k \rangle_0$, given by
\begin{equation}
P_0\langle W^k \rangle_0 = p_g e^{-\pi_g({\cal T},0)} |b_g({\cal T},0)|^2 (\hbar
\omega_0)^k + p_e e^{-\pi_e({\cal T},0)}|a_e({\cal T},0)|^2 (-\hbar \omega_0)^k.
\label{zerphot}
\end{equation}
Indeed, work $W = \hbar \omega_0$ is done by the drive whenever the two-level system
transits from $g \to e$, whereas $W = -\hbar \omega_0$ for a transition $e \to g$. The
evaluation of Eq.~(\ref{zerphot}) is straightforward. We obtain
\begin{equation}
P_0 \langle W \rangle_0 \simeq \hbar \omega_0 (p_g-p_e) (1 - \hbar \Gamma_\Sigma {\cal
T}/2) \sin^2 \lambda_0 {\cal T}/2 \hbar = \hbar \omega_0 \tanh \hbar \omega_0/2 (1 -
\hbar \Gamma_\Sigma {\cal T}/2) \sin^2 \lambda_0 {\cal T}/2 \hbar,
\end{equation}
and
\begin{equation}
P_0 \langle W^2 \rangle_0 \simeq (\hbar \omega_0)^2  (1 - \hbar \Gamma_\Sigma {\cal T}/2)
\sin^2 \lambda_0 {\cal T}/2 \hbar.
\end{equation}
As a result, the ratio
\begin{equation}
\langle W^2 \rangle_0 / \langle W \rangle_0 = \hbar \omega_0 \coth \beta \hbar \omega_0/2
\end{equation}
is independent of the strength $\lambda_0$ of the drive, {\em i.e.}, we recover the
standard fluctuation-dissipation theorem regardless of the value of $\lambda_0$, provided
we take $p_g = 1 -p_e = [1+e^{-\beta \omega_0}]^{-1}$.

Similarly, we calculate $P_0 \langle e^{-\beta W} \rangle_0$ given by
\begin{equation}
P_0 \langle e^{-\beta W} \rangle_0 = p_g e^{-\pi_g({\cal T},0)}[|a_g({\cal T},0)|^2 +
|b_g({\cal T},0)|^2 e^{-\beta \hbar \omega_0}] + p_e e^{-\pi_e({\cal T},0)}[|a_e({\cal
T},0)|^2 e^{\beta \hbar \omega_0} + |b_e({\cal T},0)|^2]. \label{zerjar}
\end{equation}
Using Eqs.~(\ref{perag}), (\ref{perae}), and (\ref{perpi}), as well as the fact that
$p_g/p_e = e^{\beta \hbar \omega_0}$ at thermal equilibrium, it is straightforward to
show that $P_0 \langle e^{-\beta W} \rangle_0 \simeq P_0$ up to corrections of order
$\Gamma ^2$. Hence $\langle e^{-\beta W} \rangle_0 \simeq 1$ in agreement with the
Jarzynski equality.

We now turn our attention to the Cayley trees of Fig.~\ref{ctree}b representing a
one-photon exchange with the environment. The total probability $P_1$ corresponding to
the eight possible quantum trajectories can be written as
\begin{eqnarray}
P_1 = \int \limits _0 ^{\cal T} dt \left\{p_g e^{-\pi_{g}(t,0)}\left[|a_g(t,0)|^2
\Gamma_\uparrow e^{-\pi_{e}(T,t)}+  |b_g(t,0)|^2 \Gamma_\downarrow e^{-\pi_{g}(T,t)}
\right]  \right.\nonumber \\ \left. + p_e e^{-\pi_{e}(t,0)}\left[|a_e(t,0)|^2
\Gamma_\uparrow e^{-\pi_{e}(T,t)}+ |b_e(t,0)|^2 \Gamma_\downarrow
e^{-\pi_{g}(T,t)}\right] \right\}.\label{P1}
\end{eqnarray}
Direct calculation shows that
\begin{equation}
P_1
 \simeq \Gamma_\Sigma {\cal T}/2 - (p_g - p_e) (\hbar
\Delta \Gamma /2\lambda_0) \sin \lambda_0 {\cal T}/\hbar,
\end{equation}
valid up to corrections of order $\Gamma^2$. Note that, as it should, $P_1 = 1 - P_0$ in
this order.

Let us look again at the moments of the work $W$ in the weak dissipation limit for the
one-photon exchange processes. We find
\begin{eqnarray}
P_1 \langle W \rangle_1 \simeq \hbar \omega_0 [\Delta \Gamma ({\cal T} /4 - ({\cal T}/8)
\cos \lambda_0 {\cal T}/\hbar - (\hbar/8 \lambda_0)\sin \lambda_0 {\cal T}/\hbar) \\
\nonumber + 2(p_g \Gamma_\downarrow - p_e \Gamma_\uparrow)({\cal T} /4 + ({\cal T}/8)
\cos \lambda_0 {\cal T}/\hbar - 3(\hbar/8 \lambda_0)\sin \lambda_0 {\cal T}/\hbar)],
\end{eqnarray}
and
\begin{eqnarray}
P_1 \langle W^2 \rangle_1 \simeq (\hbar \omega_0)^2  [\Gamma_\Sigma ({\cal T} /4 - ({\cal
T}/8) \cos \lambda_0 {\cal T}/\hbar - (\hbar/8 \lambda_0)\sin \lambda_0 {\cal T}/\hbar)
\\ \nonumber + 4(p_g \Gamma_\downarrow + p_e \Gamma_\uparrow)({\cal T} /4 + ({\cal T}/8)
\cos \lambda_0 {\cal T}/\hbar - 3(\hbar/8 \lambda_0)\sin \lambda_0 {\cal T}/\hbar)].
\end{eqnarray}
From these results we see that, different from the zero-photon case,  the ratio $\langle
W^2 \rangle_1 /\langle W \rangle_1$ generally depends on $\lambda_0$; the standard
fluctuation-dissipation relation is only recovered in the limit $\lambda_0 \to 0$, {\em
i.e.}, in the linear response regime, provided the rates $\Gamma_\uparrow$,
$\Gamma_\downarrow$ satisfy detailed balance.

Finally, we perform the calculation of $P_1\langle e^{-\beta W}\rangle _1$,
\begin{eqnarray}
P_1 \langle e^{-\beta W}\rangle_1 = \int \limits _0 ^{\cal T} dt \left\{p_g
e^{-\pi_{g}(t,0)}\left\{|a_g(t,0)|^2 \Gamma_\uparrow e^{-\pi_{e}(T,t)}[e^{\beta
\omega_0}|a_e(T,t)|^2  + |b_e(T,t)|^2]+  \right. \right.
\nonumber\\
\left. \left.+ e^{-\beta \omega_0}|b_g(t,0)|^2 \Gamma_\downarrow
e^{-\pi_{g}(T,t)}[|a_g(T,t)|^2 + e^{-\beta \omega_0}|b_g(T,t)|^2] \right\}  \right.+
\nonumber \\
\left. +p_e e^{-\pi_{e}(t,0)}\left\{e^{\beta \omega_0}|a_e(t,0)|^2 \Gamma_\uparrow
e^{-\pi_{e}(T,t)}[e^{\beta \omega_0}|a_e(T,t)|^2  + |b_e(T,t)|^2]+ \right. \right. \nonumber \\
\left. \left. + |b_e(t,0)|^2 \Gamma_\downarrow e^{-\pi_{g}(T,t)}[|a_g(T,t)|^2  +
e^{-\beta \omega_0}|b_g(T,t)|^2]\right\} \right\}\label{firjar}.
\end{eqnarray}
By direct calculation one finds $P_1\langle e^{-\beta W}\rangle _1 \simeq 1$, in
agreement with the Jarzynski equality.

Above, we analyzed the zero- and one-photon case separately. This corresponds to an
experimental situation where the photons exchanged with the environment are counted. In
experiments where the photon number remain undetermined, the statistics should be
performed on the combined zero- and one-photon processes. Clearly, the Jarzynski equality
remains satisfied, whereas the ratio $\langle W^2\rangle/\langle W\rangle$ should be
calculated as
\begin{equation}
\frac{\langle W^2\rangle}{\langle W\rangle} = \frac{P_0 \langle W^2\rangle_0 + P_1
\langle W^2\rangle_1}{P_0 \langle W\rangle_0 + P_1 \langle W\rangle_1}.
\end{equation}
It tends to the usual fluctuation-dissipation result in the linear response regime
$\lambda_0 \to 0$; deviations are found away from this limit as discussed in the main
text.

\subsection{Jarzynski equality}

Above, we have demonstrated the validity of the Jarzynski equality in the limit of weak
dissipation, using perturbation theory. We now use the Cayley tree representation of the
QJ approach to demonstrate the validity of the Jarzynski equality generally for the
driven two-level system discussed in the main text. The proof is based on certain
symmetries of the Cayley trees with respect to time reversal. Specifically, we establish
a formal relationship between the amplitudes $a(t)$ and $b(t)$ and the amplitudes
$a_R(t)$ and $b_R(t)$. The latter correspond to forward time evolution under the reversed
force protocol $\lambda_R(t) = \lambda({\cal T}-t)$, while at the same time interchanging
the role of absorption and emission such that $\Delta \Gamma \to \Delta \Gamma_R = -
\Delta \Gamma$. As a result,
\begin{eqnarray}
i \hbar \dot{a}_R = e^{-i\omega_0 t}\lambda({\cal T}-t) b_R - i \hbar \Delta \Gamma |b_R(t)|^2 a_R(t)/2, \label{dynaR} \\
i \hbar \dot{b}_R = e^{i\omega_0 t}\lambda({\cal T}-t) a_R + i\hbar \Delta \Gamma
|a_R(t)|^2 b_R(t)/2 \label{dynbR}.
\end{eqnarray}
Defining $\tilde{a}_R = e^{i \omega_0 {\cal T}/2}a_R$ and $\tilde{b}_R = e^{-i \omega_0
{\cal T}/2}b_R$, and changing the variable $ {\cal T}-t \to \tau$, we arrive at
\begin{eqnarray}
-i \hbar \dot{\tilde{a}}_R = e^{i\omega_0 \tau}\lambda(\tau) \tilde{b}_R - i \hbar \Delta \Gamma |\tilde{b}_R(t)|^2 \tilde{a}_R(t)/2, \label{dyntildeaR} \\
-i \hbar \dot{\tilde{b}}_R = e^{-i\omega_0 \tau}\lambda(\tau) \tilde{a}_R + i\hbar \Delta
\Gamma |\tilde{a}_R(t)|^2 \tilde{b}_R(t)/2 \label{dyntildebR}.
\end{eqnarray}
Comparing these equations with Eqs.~(15) and (16) from the main text, we see that the
reverse dynamics of the amplitudes $\tilde{a}_R(\tau)$ and $\tilde{b}_R(\tau)$ is
identical to the forward dynamics of $a^*(t)$ and $b^*(t)$, respectively.

Consider first the zero-photon Cayley trees shown in Fig.~{\ref{ctree}}a. Using detailed
balance, $p_e = e^{-\beta \hbar \omega_0} p_g$ , we rewrite $P_0 \langle e^{-\beta
W}\rangle_0$, Eq.~(\ref{zerjar}), eliminating the exponents $e^{\pm \beta \hbar
\omega_0}$,
\begin{equation}
P_0 \langle e^{-\beta W} \rangle_0 = e^{-\pi_g({\cal T},0)}[p_g|a_g({\cal T},0)|^2 +
p_e|b_g({\cal T},0)|^2] +  e^{-\pi_e({\cal T},0)}[p_g|a_e({\cal T},0)|^2 + p_e|b_e({\cal
T},0)|^2].
\end{equation}
Reading this result in the time-reversed direction, ${\cal T} \to 0$, the four terms
correspond to four trajectories, evolving from $g\to g$, $e \to g$, $g\to e$, and $e \to
e$, respectively. In particular, in view of the above, it corresponds to the sum of two
Cayley trees like the ones of Fig.~{\ref{ctree}}a, describing the time-evolution in
forward direction of the two-level system, driven by the reversed protocol
$\lambda_R(t)$, and upon exchanging the role of absorption and emission. Indeed,
Eqs.~(\ref{dyntildeaR}) and (\ref{dyntildebR}) imply that
\begin{eqnarray}
|a_g(t_1,t_2)|^2 e^{-\pi_{g}(t_1,t_2)} &=& |a_{R,g}(T-t_2,T-t_1)|^2 e^{-\pi_{R,g}(T-t_2,T-t_1)}, \label{rel1}\\
|b_g(t_1,t_2)|^2 e^{-\pi_{g}(t_1,t_2)} &=& |a_{R,e}(T-t_2,T-t_1)|^2
e^{-\pi_{R,e}(T-t_2,T-t_1)}, \\ \label{rel2}
|a_e(t_1,t_2)|^2 e^{-\pi_{e}(t_1,t_2)} &=& |b_{R,g}(T-t_2,T-t_1)|^2 e^{-\pi_{R,g}(T-t_2,T-t_1)}, \label{rel3}\\
|b_e(t_1,t_2)|^2 e^{-\pi_{e}(t_1,t_2)} &=& |b_{R,e}(T-t_2,T-t_1)|^2
e^{-\pi_{R,e}(T-t_2,T-t_1)}.  \label{rel4}
\end{eqnarray}
Hence, $P_0 \langle e^{-\beta W} \rangle_0 = P_{R,0}$, the probability for the system to
evolve according to the reverse protocol without exchanging a photon with the
environment.

Similarly, one can analyze the one-photon Cayley trees shown in Fig.~{\ref{ctree}}b.
Using detailed balance both for $p_{g,e}$ and for $\Gamma_{\downarrow,\uparrow}$, we
rewrite Eq.~(\ref{firjar}) as
\begin{eqnarray}
P_1 \langle e^{-\beta W}\rangle_1 = \int \limits _0 ^{\cal T} dt \left\{
e^{-\pi_{g}(t,0)}\left\{|a_g(t,0)|^2  \Gamma_\downarrow e^{-\pi_{e}(T,t)}[|a_e(T,t)|^2
p_g + |b_e(T,t)|^2 p_e]+ \right. \right.
\nonumber\\
\left. \left.+ |b_g(t,0)|^2 \Gamma_\uparrow e^{-\pi_{g}(T,t)}[|a_g(T,t)|^2 p_g +
|b_g(T,t)|^2p_e] \right\}  \right.+
\nonumber \\
\left. +e^{-\pi_{e}(t,0)}\left\{|a_e(t,0)|^2 \Gamma_\downarrow
e^{-\pi_{e}(T,t)}[|a_e(T,t)|^2 p_g + |b_e(T,t)|^2] p_e+ \right. \right. \nonumber \\
\left. \left. + |b_e(t,0)|^2 \Gamma_\uparrow e^{-\pi_{g}(T,t)}[|a_g(T,t)|^2 p_g  +
|b_g(T,t)|^2 p_e]\right\} \right\}.
\end{eqnarray}
With the help of Eqs.~(\ref{rel1}) -- (\ref{rel4}), this reads
\begin{eqnarray}
P_1 \langle e^{-\beta W}\rangle_1 =\nonumber \\
= \int \limits _0 ^{\cal T} dt \left\{
e^{-\pi_{R,g}({\cal T},{\cal T}-t)}|a_{R,g}({\cal T},{\cal T}-t)|^2 \Gamma_\downarrow
[e^{-\pi_{R,g}({\cal T} -t,0)}|b_{R,g}({\cal T} -t,0)|^2 p_g + e^{-\pi_{R,e}({\cal T}
-t,0)}|b_{R,e}({\cal T} -t,0)|^2 p_e]+ \right.
\nonumber\\
\left. + e^{-\pi_{R,e}({\cal T},{\cal T}-t)} |a_{R,e}({\cal T},{\cal T}-t)|^2
\Gamma_\uparrow [e^{-\pi_{R,g}({\cal T} -t,0)}|a_{R,g}({\cal T} -t,0)|^2 p_g +
e^{-\pi_{R,e}({\cal T} -t,0)} |a_{R,e}({\cal T} -t,0)|^2p_e]   + \right.
\nonumber \\
\left.
 +e^{-\pi_{R,g}({\cal T},{\cal T}-t)}|b_{R,g}({\cal T},{\cal T}-t)|^2
\Gamma_\downarrow [e^{-\pi_{R,g}({\cal T} -t,0)}|b_{R,g}({\cal T} -t,0)|^2 p_g +
e^{-\pi_{R,e}({\cal T} -t,0)}|b_{R,e}({\cal T} -t,0)|^2 p_e
+ \right. \nonumber \\
\left. + e^{-\pi_{R,e}({\cal T},{\cal T}-t)}|b_{R,e}({\cal T},{\cal T}-t)|^2
\Gamma_\uparrow [e^{-\pi_{R,g}({\cal T} -t,0)}|a_{R,g}({\cal T} -t,0)|^2 p_g +
e^{-\pi_{R,e}({\cal T} -t,0)} |a_{R,e}({\cal T} -t,0)|^2 p_e]\right\}.
\end{eqnarray}
Regrouping terms and changing the integration variable ${\cal T} - t \to \tau$, this
expression can be shown to be of the form of Eq.~(\ref{P1}). It thus represents the
probability $P_{1,R}$ corresponding to all the trajectories for one-photon processes to
occur under the reversed protocol.

The extension to the analysis of $n$-photon processes is straightforward, with the
obvious result $P_n \langle e^{-\beta W}\rangle_n = P_{R,n}$.  Summing over $n$ on both
sides yields Jarzynski's equality $\langle e^{-\beta W}\rangle = 1$.

\end{widetext}

\begin{thebibliography}{99}
\bibitem{footnote} Throughout this paper, we will work in the limit where the conditions for the usual Bloch-Redfield approximation (small relaxation rates and slow response of the two-level system as compared to the bath's correlation frequency) are satisfied.
\end{thebibliography}
\end{document}